# Identifying Missing Component in the Bechdel Test Using Principal Component Analysis Method

Raghav Lakhotia, Chandra Kanth Nagesh, Krishna Madgula

***Abstract*—**A lot has been said and discussed regarding the rationale and significance of the Bechdel Score. It became a digital sensation in 2013, when Swedish cinemas began to showcase the Bechdel test score of a film alongside its rating. The test has drawn criticism from experts and the film fraternity regarding its use to rate the female presence in a movie. The pundits believe that the score is too simplified and the underlying criteria of a film to pass the test must include 1) at least two women, 2) who have at least one dialogue, 3) about something other than a man, is egregious. In this research, we have considered a few more parameters which highlight how we represent females in film, like the number of female dialogues in a movie, dialogue genre, and part of speech tags in the dialogue. The parameters were missing in the existing criteria to calculate the Bechdel score. The research aims to analyze 342 movies scripts to test a hypothesis if these extra parameters, above with the current Bechdel criteria, are significant in calculating the female representation score. The result of the Principal Component Analysis method concludes that the female dialogue content is a key component and should be considered while measuring the representation of women in a work of fiction.

## I. INTRODUCTION

DIVERSITY and inclusion are the two words that have gained prominence in our daily conversations. There are extensive debates and discussions on equal rights for men and women. The current generation is now realizing the disturbing gender gap and wants to express their views so that they do not miss being a part of this revolution. Social media has been trending with such posts and articles. Some credit goes to the #MeToo movement as well. The virtual world is trying to make a point viz. we under-represent women across all industries and domains, and the math proves it. As per 2018 data from United Nations [1], we see women are highly underrepresented in senior management positions. The World Bank published a report showcasing the firms which have a top female manager i.e. CEO or equivalent highest ranking position [2]. As per the report, only 19% of the firms are being managed by a female CEO and interestingly these 19% firms tend also to be firms with more female workers.

A similar situation persists in the entertainment industry as well. Numerous reports have revealed depressing statistics about how women are under-represented on screen. According to the report "Celluloid Ceiling" [3], "women comprised only 18% of all the directors, writers, producers, editors, executive and cinematographers working on the top 250 domestic grossing films in 2017". While in 2016, only 12% of films out of 900 had a balanced cast [4]. Men outnumber women in the film industry as well. The issue has long prevailed in the industry since, but the question arises; did we highlight this gender bias to the extent that we should have?

In 1985, Alison Bechdel, an American cartoonist, invented a score [5] to gauge gender equality on screen. The score thrives on three rules; a) the film should have two named female characters, b) the characters should have at least one dialogue among them, and c) the speech should be something other than a man. On applying this score over the movies featured in Oscar history, we see only 36 out of 89 the best movie winners could clear this score. The score has become a touchstone for talking about feminism and film. In November 2013, Swedish cinema introduced the Bechdel Score as a parameter for movie ratings to highlight gender bias [6]. As per the cinema, a movie needs to have high Bechdel score i.e. pass all three criteria to get an "A" rating. The move has been supported by a few but has been criticized by many as well. Critics and researchers feel that the Bechdel score is a blunt tool that does not reveal if the movie is gender-balanced. The rules are so simple that a film can pass the criteria with a single dialogue between two female characters where they discuss their shopping. But for a movie like Gravity, which has a female as the main protagonists, the test fails.

So, is Bechdel score the right lens through which to discuss gender equality? Are there some limitations to the score? Is the score suited only for multi-character films? After Bechdel, possible attempts were made to get that perfect score to gauge gender equality on the screen. Hence, new tests arose. In 2016, [7] introduced the DuVernay Test, where the involvement of African-American or other minorities was an essential criterion for female representation. Still, female representation behind the cameras was ignored. Hence in 2017, "The Next Bechdel Test" was proposed by [8], where they took into account the female representation behind the camera as one of the main criteria for a movie to pass or fail on the gender equality KPI. According to the test, if the on-set crew had 50% or more women, then the movie scores high on female representation.

## II. MOTIVATION

We see numerous tests were invented and used as a tool for calculating gender equality in the film industry. But did we do

Raghav Lakhotia is currently working with General Electric India Pvt Ltd. Bangalore, India (corresponding author, phone: +91-9591192190; e-mail: Raghav.lakhotia.2015@iimu.ac.in).

Chandra Kanth N and Krishna Madgula are currently working with General Electric India Pvt Ltd. Bangalore, India (e-mail: canfindck@gmail.com, krishnac.madgula@gmail.com).

justice in labeling the film as "sexist" based on a single dialogue, or the number of female members present in the movie. We need to dig a step deeper and see how the female is represented on screen and how their character is portrayed in the movie. In this paper, the authors try to put forward a comprehensive approach to calculate the on-screen female representation score. To achieve such a score, we believe it is important to learn how female characters are presented on screen. Female representation should not be restricted to the number of female characters or the number of dialogues delivered. Instead, it should analyze the emotions or sentiments in the communications produced by the female character in the movie. There can be cases when there is one female character, who is in the lead role and delivers maximum dialogue based on the movie theme. Or there can be a case when there are multiple female characters with a cameo role, and each has conversations that have no relation to the subject of the movie. In both the cases, the movie will fail all the present/proposed tests on gender equality. Key factors taken into account in the research, apart from Bechdel Tests, for both male and female are – a genre of the dialogues spoken, parts of speech (POS) tags i.e. the number of adjective, verbs and nouns and dialogues sentiments. Through the Principal Component Analysis (PCA) method, we aim to reduce the complexity over multiple variables and try to highlight the most significant parameters which should be used to calculate the female representation score.

## III. METHODOLOGY

The work aims to highlight if the parameters like female dialogue genre, female dialogue POS tags have a significance over and above the existing Bechdel score in calculating a comprehensive score for female representation on-screen. Hence, the hypothesis is as follows:

H1: The parameters like female dialogue genre, POS tags and sentiments are significant in highlighting the female representation on-screen.

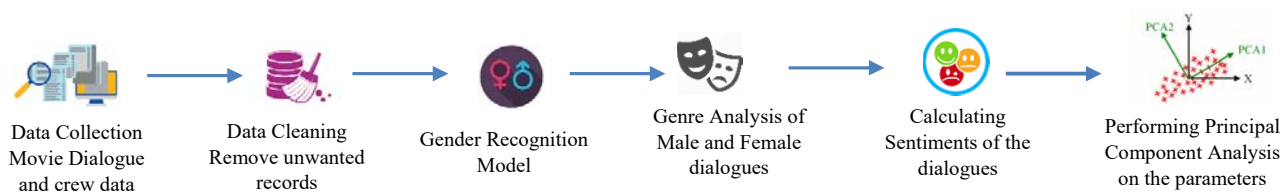

Fig. 1 Steps involved in our research

Evaluation of the movies on female representation scale was done using Principal Component Analysis (PCA). The research takes into consideration ten items/parameters that measure the female representation On-screen and Off-screen regarding various aspects of their role/participation in making the film. The purpose of PCA is to reduce the number of variables that measure the female representation in a movie and also ensure that the variables are not correlated.

## IV. DATA SET

We used the Cornell Dataset [9] for 617 movies under the tag of Cornell Movie Dialogue Corpus [9]. The corpus contains a large metadata-rich collection of conversations extracted from raw movie scripts. It involves 9,035 characters from 617 movies having 220,579 conversational exchanges between 10,292 pairs of movie characters. This data were later merged with the Kaggle's - The Movies Dataset [10], to gather the crew, budget, revenue and ROI for each of the movies. Finally, we considered 324 movies for our research because of unavailability of data across all parameters out of 617 movies.

The corpus is divided into five sections:

a) Movie Metadata – The file contains information about each movie title with fields viz. movie ID, title, release year, IMDB rating, number of IMDB votes and genre. A movie always falls into one or more genre.

b) Character Metadata – The file contains information about the characters in each movie. A unique character ID represents each character in the dataset. The character ID is tagged to the respective movie ID. The sheet also contains the gender details (M/F) for 3,774 characters out of 10,741. For the remaining 6,979 characters, gender analysis model was used to classify the gender information. Subsequent sections explain the model in detail.

c) Movie Dialogue – The file contains the actual text of each dialogue or utterance in the movie data set. Each dialogue is represented by a line ID, tagged to a particular character of the respective movie. For example, the dialogue from a movie is represented by the line ID, the person who spoke the dialogue (character), the character id and the movie name.

d) Movie Conversations - The file defines the structure of the conversations between the two characters. Each record depicts the two characters involved (tagged through character ID) in the discussion in a particular movie (labeled through movie ID). The file is structured in the following way.

The list of utterances is the collection of movie dialogues spoken by a character from the movie. It translates to a group of line ID and is stored in a particular order which signifies the line number of each dialogue spoken by the character in the script. Even the character ID of whom they talked to is also mentioned in the meta-data. For example, character $U_0$ speaks to character $U_2$ and the list of utterances is denoted by ['L194', 'L195', 'L196', 'L197'], which corresponds to the line number of actual dialogues. Each record is tagged with a unique movie

ID.

e) Crew Metadata - The file contains the crew information for all the movies. The crew details included the crew member's name, his/her job title, which can range from the director of the movie to the key hair stylist, and their gender. The dataset also contains the budget and revenue. The ROI is calculated using the following formulae:

$$ROI = \left[\frac{(Revenue - Budget)}{Budget}\right] X\ 100 \quad (1)$$

## V. Pre-Processing and Data Cleaning

Although the Cornel data set was clean, still, some work was required to fill in the incomplete data like the gender of the characters in the data set. Out of 10,741 characters, the data set held gender information for only 3,774 characters. Machine learning techniques called N-grams was used to detect the gender of the characters through their names. In this section, we will discuss the different methods/models used to complete the missing values in the data set.

### A. Gender Recognition Model

N-gram is an n character slice or sequence of a longer string. The notation helps to capture the language structure like which letter or word is likely to follow the given letter or a word in a sentence. In our system, we use N-grams of different lengths, which range from one to three character lengths. These different character length grams are usually referred to as uni-grams, bi-grams and tri-grams. We consider the character sequence N-grams from the reverse order of a name. In this way, we capture the leading and the trailing features and use that to model the decision tree classifier. Thus, a simple example of a name "MARY JANE" would be composed of the following N-grams:

**uni-grams**: M, E
**bi-grams**: MA, NE
**tri-grams**: MAR, ANE

The gender recognition procedure is as follows:

### 1) Obtain Training Set

We performed our experiments on the data set namely – "*Popularity of the number of baby names*" over the years – published by the US Social Security Administration [11]. The data set contains 34,425 boy's names and 60,600 girl's names for a total of 95,025 baby names. Each record in the dataset has the name and the gender associated with it. We did not consider records with an unknown gender value. Later, we considered 90% of the data as the training set and the remaining 10% data as the validating set.

### 2) Compute Dataset for N-Gram

For the existing data set, we created uni-gram, bi-gram and tri-gram representations of each record using basic string operations. As discussed earlier, we considered the first letter, last letter, first two letters, last two letters, first three letters and last three letters as our N-grams. These feature sets are later fed into a vectorizer. For the vectorizer, we use the Scikit-Learn library [12], and it provides vectors based on the feature-value mapping dictionary. Feature-value mapping is nothing but uni-grams, bi-grams and tri-grams as explained earlier [13]. Then, we cross-validate our dataset in order to maintain a mixed and uneven data distribution so that we get a generalized feature learning set for our Machine Learning model.

### 3) Classifier Model

Since our target variable (in our case is the gender) can take up a discrete set of values, we model a Decision Tree Classifier to predict the gender using the variable "*Name*" as the input. In tree structures, we learn that the leaves represent the classes or class labels, whereas the nodes represent the features that lead to the class labels. To model the decision tree, we use the Scikit-learn library [12]. Scikit-learn is the most widely used and famous Python machine learning system. It contains a diversified set of libraries and utilities for many of the machine learning models.

In performing any machine learning tasks, words cannot be used directly as they are considered categorical variables. Categorical variables often tend to be broken up into values that are names or labels, and there is no intrinsic ordering to the categories. For example, gender is a categorical variable, with two values either a male or a female. Most machine learning models take numerical values as inputs. Therefore, for the machine learning model to predict a categorical variable, we first need to convert the names into a number sequence through the use of the vectorizer method. The main job is to convert the words into word vectors. A vectorizer converts a collection of texts into a matrix of token counts. It learns what are the essential words and their features. Fig. 2 gives a snapshot of a single node of the classifier model. The figure illustrates the decision tree with its nodes and leaves and the classification rules. For example, if the first-letter is j and is true, then it checks the first three letters of the same word. If it is jac, it classifies it as a female, else continues to further classify the name with respect to the various features.

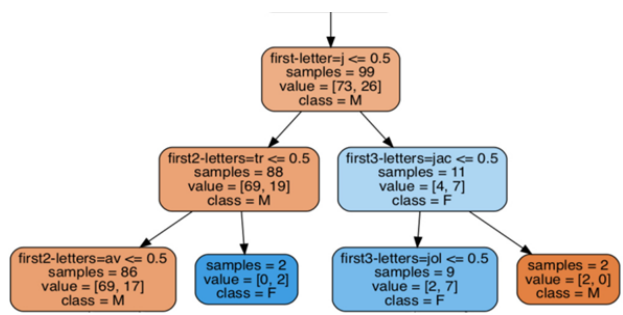

Fig. 2 Decision tree of our gender classifier model

This decision tree classifier yielded an accuracy score of 98.67% on the training set and 86.9% on the testing set. After identifying the gender of the missing characters in our data set, we found that the gender ratio in the movies across years is dismal. Fig. 3 shows the movie distribution across the year and the gender ration for cast and crew in the movies. For every three male, there is one female character present in the cast and crew.

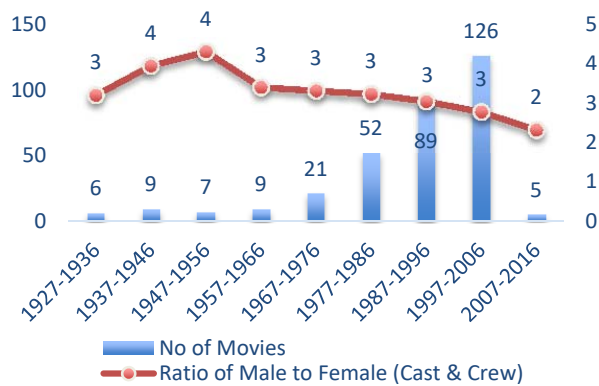

Fig. 3 Movie count and gender ratio in cast & crew in the data set

### B. Genre Tagging

Film genre is the various forms, categories, classification or groups of the film based on a specific recurring pattern, style, theme or syntax. For example, a movie which falls under the genre action will have high energy showcased in the form of battles and fights, like Godzilla and X-men. While a romantic film will have a majority of narratives or scenes depicting love, and exchanging emotions etc., like Eternal Sunshine of the Spotless Mind and Rocky. In the sample data collected from the Cornell dataset, movie genres were stated. Each movie will have multiple genre. For our simplicity, we took the major three genres of the film.

In this activity, we aim to find the genre of the dialogues delivered by the male and female characters ($G_m$, $G_f$, respectively) through an approach which involves finding out the cosine similarity of each dialogues to the genre of the movie. The process will be explained in more detail in the coming sections. Once identified, we calculate the distance between $G_m$, $G_f$ w.r.t movie genre ($G_{mo}$). If the distance is high, it indicates the dialogues delivered by either male or female have more significance w.r.t the overall genre of the movie and vice versa.

Dialogue plays a crucial role in establishing the logic of the film sequence. The primary motive of the dialogue is to communicate "why", "how" and "what next" in the movie. The dialogue paves the way for the viewers to understand and interpret what is shown in a particular scene [14]. Things like a sense of urgency cannot be explained visually. The dialogue also helps the audience to understand the character's personality. It is not always the way of speaking where the dialect or accent gives the audience clues about the character but also what is said by the character. Apart from the dialogue delivered by the person itself, relevant evidence is given on the type of character played by a person in the movie through dialogue delivered by other people as well. Hence, we feel, it is imperative to identify the style or genre of the dialogue delivered by each character to understand the message conveyed in context with the film genre. It will also help us to know how the character is being portrayed in the movie, whether significant to the movie genre or vice versa.

To find the genre of the dialogue, we used word embedding. Word embedding is referred to as numerical representation of words. We need to convert string into numbers which is fed as the input to perform classification or any other job. A lot of applications have been built using word embedding techniques e.g. document classification, news classification, sentiment analysis etc. [15], [16].

The first step is to pre-process each dialogue. We start the pre-processing by tokenizing the entire dialogue and removing the stop words. Stop words are regularly or frequently used common words which occur in every sentence. These include words like he, she, him, the, her, who etc. After removing the stop-words, we get a more clear and robust dialogue which now focuses mainly on the context and the key words spoken in the dialogue. We use NLTK's stop-words dataset to remove the stop-words from the dialogues. Then we convert the stripped dialogues into their word embedding. For this we use Word2Vec models.

Word2Vec is a computationally efficient predictive model for learning word embedding from raw text. It is composed of two flavors - the Continuous Bag-of-Words called (CBOW) and the Skip-Gram model. In a way, both these models are similar, except that CBOW predicts target words (e.g. 'ball') from source context words ('the player catches the'), while the skip-gram does the inverse and predicts source context-words from the target words. Word embedding model can be divided into two categories - Count based methods and Predictive methods. Models in both categories share the assumption that words that appear in the same contexts share semantic meaning. Predictive methods try to predict a word from its neighbors in terms of learned small, dense embedding vectors [17]. We use Word2Vec method to generate word embedding for our work.

For our work, we used Google's pre trained Word2Vec model that was trained on Google News Dataset, and the model contains 300-dimensional word vectors called word-embeddings for over 3 million words [18]. Suppose a movie, Mj, is categorized by genres $G_j = \{g_1, g_2, \ldots \ldots, g_k\}$. Let $d = \{w_1, w_2, \ldots \ldots, w_n\}$ be any particular dialog in the movie, where $w_i$ corresponds to the word i. The vector representation of the dialog is given by:

$$Dialog\ Embedding\ (D) = \sum_{i=0}^{n} Word\ Embedding\ (W_i) \big/ n \quad (2)$$

where, $Word\ Embedding\ (W_i)$ represents the word embedding of the individual word.

Similarly, we obtain the word embedding for each genre category of the movie represented as G. The simplest property of embedding obtained from above method is that similar words tend to have similar vectors. For example, the words king is to queen as the words male is to female. There is a coherency and similarity between the words. In mathematical terms, the similarity between two words (as rated by humans on scale of [-1,1]) correlates with the cosine similarity between those word vectors. Later we compute the cosine distance between each dialog from the Dialog embeddings (D) and the movie genre (G). We pick the top three genre of each movies and calculate the cosine similarity between them. Suppose that we have the dialog embeddings $D_j = \{D_1, D_2, \ldots\ldots, D_k\}$ and the genre embedding $G = \{g_1, g_2, g_3\}$. We calculate the cosine similarity between both these vectors which is represented as the dot product:

$$Cosine\ Similarity\ (g_t, D) = \frac{D.g_t}{|D|.|g_t|} \qquad (3)$$

where D – Stands for Dialog Embedding, $g_t$ – Stands for Word Embedding, and $1 \le t \le 3$.

These cosine similarity measures are stored and sorted with respect to females and males.

$$Cosine\ Similarity\ Male\ (CS_m) = \Sigma\ (CS_m\ (g_1, D_m) + CS_m\ (g_2, D_m) + CS_m\ (g_3, D_m)) \qquad (4)$$

$$Cosine\ Similarity\ Female\ (CS_f) = \Sigma\ (CS_f\ (g_1, D_f) + CS_f\ (g_2, D_f) + CS_f\ (g_3, D_f)) \qquad (5)$$

where, $D_m$ is a male dialogue, $D_f$ is a female dialogue.

This value is later averaged by dividing the corresponding score with the number of dialogues of male and female in the movie.

$$Genre\ Tag\ Male\ (G_m) = \frac{CS_m}{\sum Male\ Dialogues} \qquad (6)$$

$$Genre\ Tag\ Female\ (G_f) = \frac{CS_f}{\sum Female\ Dialogues} \qquad (7)$$

### C. Part of Speech (POS) Tagging

Dialogue is a conversation between two or more characters in a work of fiction. It helps the writer to reveal character traits, and it contains the character's emotions and presents their personality in a scene. Dialogues are made of words which are tagged to one of the parts of speech (POS) in the English language. The parts of speech are the adjective, noun, verb etc. We aim to analyze the trend of POS tags for the dialogues delivered by male and female characters across our data set which includes 324 movies. We used the NLTK pre-trained supervised model to compute the POS tags in each dialogue. Maximum Entropy classifier is one such default pre-trained model in NLTK. The model was trained on the Wall Street Journal Corpus.

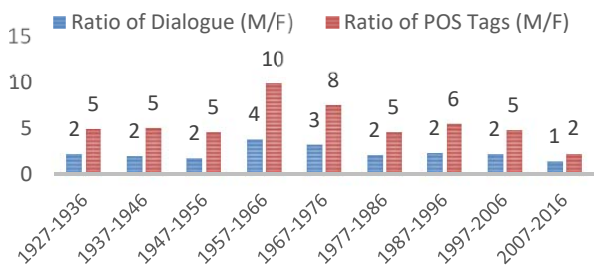

Fig. 4 Male to Female Dialogue Ratio and POS Tag Ratio in the data set

Fig. 4 highlights the ratio of dialogues and the POS tags in those dialogues delivered by male as compared to female characters in the movies across the studied years for our data set. For the majority of the years, male characters spoke twice the times of dialogue spoken by their female counterparts. Also, the dialogue of males has five times the POS tags compared to the dialogue of females. The figure clearly demonstrates the lack of parity in the dialogues assigned to a male character in a movie as compared to a female character.

### D. Sentiment Analysis

Dialogue sentiment represents the emotion and mood of the transcript exchanged between two or more characters in a movie. Decoding and learning the sentiments of a dialogue also helps to understand the character played by a person in a movie. Statistically, sentiment analysis mainly represents whether a dialogue is positive, negative or neutral. We have used NLTK's VADER sentiment analysis package [19], where VADER stands for Valence Aware Dictionary and Sentiment Reasoned, to capture the dialogue sentiment in the movie scripts. VADER sentiment analysis belongs to a class of lexicon and rules based on sentiment-related words which are more attuned to the words from social media [19]. The model is sensitive to both polarity (positive/negative) and intensity (strength) of emotion in a dialogue. Lexical approach identifies the sentiment of a word by building a lexicon i.e., dictionary of sentiment, and mapping words to it. In this approach, we are not required to train the model using labeled data. The scores of a word in the VADER lexicons library ranges from -4 to 4 [20], where -4 is the most negative and +4 is the most positive. For example, the dialogue "You're concentrating awfully hard considering it's gym class!" has two words in the lexicon (awfully and hard) with ratings of -2.0 and -0.4, respectively. The composite score is calculated by summing up the sentiment score of each VADER dictionary listed word in a sentence. The returned score ranges from -1 to 1. The normalization is done using the formula:

$$Final\ Score = \frac{x}{\sqrt{x^2 + a}} \qquad (8)$$

where, x is the sum of the scores of the constituent words in a sentence and $a$ is the normalization parameter which is set to default of 15.

VADER also considers various heuristics like punctuations, and use of capital words, which imparts intensity to the emotion. If all the words are capital, it adds or decreases a particular value in the sentiment of the word depending on whether the word is positive or negative.

## VI. THE PRINCIPAL COMPONENT ANALYSIS METHOD (PCA)

Principal Component Analysis [21] is a well-known technique for dimensionality reduction and multivariate analysis. It is used as a major tool across various fields like data compression, pattern recognition, time series analysis and image processing. A complete discussion of PCA methodology can be found in textbooks [22], [23]. PCA is a statistical multivariate technique that uses orthogonal transformation to convert several correlated observed variables into a smaller number of uncorrelated variables known as principal components. The main objective of the unsupervised learning method are to reduce dimensionality, scoring all observations based on a composite index and clustering observations together which show similar multivariate attributes. Reducing a $d$ multivariate attributes by two or three variables helps in displaying the results graphically which is easy to understand.

Multivariate analysis is a technique which uses data sets that have more than one response variable for each observational unit. The data set can be summarized as a data matrix $X$ having $n$ rows and $p$ columns. The rows represents the observations and columns denotes the variables in the model. In deriving the Principal Component (PC), the correlation matrix is used instead of covariance matrix as different variables in the data set are measured using different units and have different variances. Using correlation matrix is equivalent to standardizing the variable to zero mean and unit standard deviation. The PCA model can be represented as:

$$U_{mx1} = W_{mxd}X_{dx1} \qquad (9)$$

where U, an m-dimensional vector is a projection of $X$ – the original $d$-dimensional data vector (m<<d).

The $m$ projections that maximizes the variance of $U$, called the principal axes, are represented by the Eigenvectors $e_1$, $e_2$, $e_3$, …, $e_m$ of the data set correlation matrix $S$. The data set correlation matrix $S$ can be found through:

$$S = \frac{1}{(n-1)}\sum_{i=1}^{n}(X_i - \mu)(X_i - \mu)^T \qquad (10)$$

where $\mu$ is the mean vector of $X_i$.

The Eigenvalues $e_i$ can be found through solving set of equations:

$$(S - \lambda_i I)e_i = 0, i = 1,2,…,d \qquad (11)$$

where $\lambda_i$ are the eigenvalues of S. The eigenvectors are sorted based on the magnitude of the corresponding eigenvalues. The eigenvalue represents the level of variation caused by the associated principal component. The variance for the principal component for k-retained principal components is computed by:

$$t_k = \frac{\sum_{i=1}^{k}\lambda_i}{\sum_{i=1}^{d}\lambda_i} \qquad (12)$$

## VII. Result

The Kaiser-Meyer-Olkin (KMO) measure of sampling adequacy and Bartlett's sphericity test (BTS) results are presented in Table I.

According to [24], if the KMO value is greater than 0.5 and the BTS less than 0.05, the data is suitable for PCA. In this study, the KMO coefficient of 0.69 is adequate and the Bartlett's test is significant at 99% (p < 0.005).

The heat-map in Fig. 5 presents a snapshot on the correlation between the parameters used in the analysis. As per the observation, strong correlation can be seen between the POS tags of each of the dialogues. Apart from that, there were no significant correlations among various other parameters. Thus it indicates that the data are very well spread out and the randomness in the data is quite high.

TABLE I
KAISER-MEYER-OLKIN (KMO) AND BARTLETT'S TEST

| KMO and Bartlett's Test | | |
|---|---|---|
| Kaiser-Meyer-Olkin Measure of Sampling Adequacy | | 0.698 |
| Bartlett's Test of Sphericity | Approx. Chi Square value | 1210.54 |
| | Degrees of Freedom | 4 |
| | Significant Value | 0 |

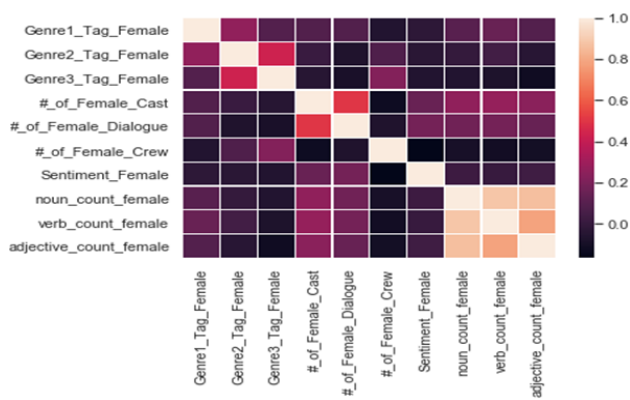

Fig. 5 Heatmap between various features of the data-frame

In order to perform PCA, we obtained the eigenvalue decomposition matrix from the correlation matrix. The eigenvalues explain the variance of the data along the feature axis. The decomposition matrix is projected on the principal axes and the projections are called Principal Components or the PC scores. The eigenvalues are arranged in the descending order of their magnitude to provide a ranking of the components or axes. The eigenvalues close to and less than zero represent components or axis which can be discarded and vice-versa. Table II shows the PC eigenvalues and percentage of variance explained.

TABLE II
Principal Components with Variance

| | Eigen Values | % of Variance | Cumulative % |
|---|---|---|---|
| **PC1** | 2.970 | 29.703 | **29.703** |
| **PC2** | 1.641 | 16.406 | **46.109** |
| **PC3** | 1.406 | 14.059 | **60.168** |
| **PC4** | 1.034 | 10.344 | **70.513** |
| PC5 | 0.915 | 9.153 | 79.665 |
| PC6 | 0.734 | 7.337 | 87.002 |
| PC7 | 0.522 | 5.223 | 92.224 |
| PC8 | 0.474 | 4.744 | 96.968 |
| PC9 | 0.210 | 2.103 | 99.071 |
| PC10 | 0.093 | 0.929 | 100.000 |

The first four components have eigenvalues greater than 1 and have cumulative variance of 70% (bold and highlighted in yellow). Hence, we take these components for further analysis.

There is another method to identify the important components to be considered for the analysis. This method is a bit subjective in nature but we used it to cross-verify our results obtained above. According to the method, the significant factors are disposed like a cliff, having a huge and evident slope, while the less significant factors usually have negligible slope.

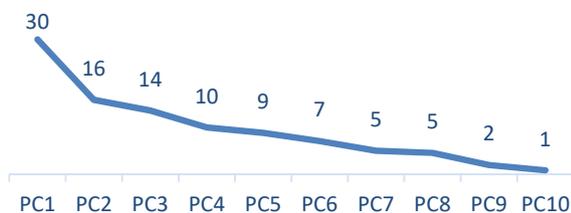

Fig. 6 Scree Plot for initial variables with % variance

In Fig. 6, we can see that the initial four components have huge slope and after the fourth component, the slope of the curves is very small. So these factors should be eliminated from the model. Hence from both the models, we conclude that the four components account for 70% of the total variance should be used for factor loading.

After identifying the principal components, we determine the factor loading or the component loading. Factor loading is the correlation coefficients between the variables (rows) and factors (columns). The mathematical formula is:

$$Factor\ Loading = eigen\ vector\ X\ \sqrt{eigen\ values} \quad (13)$$

TABLE III
FACTOR LOADINGS OF EACH COMPONENT

|  | PC1 | PC2 | PC3 | PC4 |
|---|---|---|---|---|
| **Genre1 Tag Female** | -0.192 | **-0.400** | -0.338 | -0.467 |
| **Genre2 Tag Female** | 0.033 | **-0.741** | -0.340 | -0.191 |
| **Genre3 Tag Female** | 0.170 | **-0.716** | -0.244 | 0.171 |
| **Number of Female Cast** | -0.493 | 0.169 | **-0.590** | 0.270 |
| **Number of Female Dialogues** | -0.392 | 0.271 | **-0.628** | 0.345 |
| **Number of Female Crew** | 0.220 | -0.383 | 0.140 | **0.721** |
| **Sentiment of Female** | -0.122 | 0.327 | -0.435 | -0.197 |
| **Noun counts of Female** | **-0.919** | -0.155 | 0.240 | 0.030 |
| **Verb counts of Female** | **-0.902** | -0.174 | 0.197 | -0.007 |
| **Adjective counts of Female** | **-0.886** | -0.095 | 0.264 | 0.006 |

The variables that have high loading (absolute value) on a factor are considered significant in terms of their similarity regarding the measured construct. The negative sign behind the values signify the direction in the d-dimensional space. Once done for all the variables across each principal component, the component is labeled according to the relevant meaning. In Table III, we can see that the first component is determined by the variables labeled as Noun Counts, Verb Counts and Adjective Count having high absolute factor loading (bold and highlighted in yellow). The second principal component is determined by variables Genre 1 tag, Genre 2 tag and Genre 3 tag of the female dialogues and the third principal component by number of female cast and dialogues delivered by them in a movie (bold and highlighted in yellow).

The fourth principal component has only one significant variable namely number of female members in the crew. Taking into account the above observations, we can label the first PC as "*Content of the female dialogue*", the second principal component as "*Female Dialogue Category/Genre*" and the third principal component as "*Gender Diversity in the Cast and Crew*". Thus, we accept the alternate hypothesis saying the parameters like female dialogue genre, POS tags and sentiments are significant in highlighting the female representation on-screen.

## VIII. CONCLUSION

PCA is a powerful tool to decide the best factor that describes the total variance produced by the primary variables. In such cases, when we use multiple variables which demonstrates multi-collinearity behaviour among themselves, then another method like regression analysis is not the correct method to determine the significant factors. Analysis of PCA on the data set shows that the female representation should be indicated by three components viz. the *Content of the female dialogue*, *Category/Genre of the Female dialogue w.r.t genre of the movie* and the *gender diversity in the cast & crew of the movie*. As per the research, no component is in accordance with the existing criteria of the Bechdel test. The current Bechdel test criteria saying – a female should have at-least one line of dialogue and not talk about a man, has no resemblance with the research finding to the component – Category/Genre of the female dialogue or the Content of the female dialogue. Instead, we propose that even though the female conversations might revolve around a male character, it is significant if it aligns with the genre of the movie. Gender diversity in cast and crew as one of the main findings of our research finding is in following with the conclusions proposed by [8]. Thus, using scientific models and sufficient proof, we recommend that a movie should be rated as high or low on female representation based on the content of the female dialogues and the theme/category of those dialogues concerning the overall genre of the movie. The film directors and story writers should include more female characters in the cast and crew of the movie, should assign dialogues to them which aligns with the genre or the central theme of the movie. These criteria, if considered, will result in producing balanced movies having high female representation. In our future work, we aim to capture the on-screen time spent by the female characters in the movie and analyze if this variable has a significant impact on the female representation score in a film.

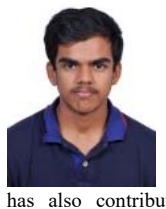

**Madgula Krishna Chaitanya** is currently working for GE Power as a Data Engineering Specialist. He has completed his Bachelors in Computer Science and Engineering from PES Institute Of Technology, Bangalore South Campus over the period 2014-2018. He has worked on quite on a few academic research projects in the fields of Machine Learning, Deep Learning, Natural Language Processing and Astro-informatics and has also contributed to a research paper published in Association for Computational Linguistics. He actively participates in Data Science hackathons and has also been a data science mentor for a college hackathon. He has also presented a technical paper at a national level conference. He is very enthusiastic about ideation and implementation of technological ideas He can be reached out at **e-mail id**: krishnac.madgula@gmail.com.

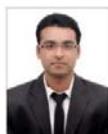

**Raghav Lakhotia** is currently working in GE, as a part of Digital Technology Leadership Program. He completed his MBA from Indian Institute of Management Udaipur, 2015-17. Prior to his MBA, he has 3 years of work experience in IT domain as a developer and researcher. He has worked two years in Samsung and one year in Verizon. During his stint in Samsung, he has published five technical papers in IEEE and various International Science Journals. He loves to ideate on technical topics or general research domains. In his free time, he loves to play cricket. He can be reached out at **e-mail id** – raghav.lakhotia.2015@iimu.ac.in.

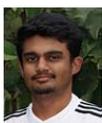

**Chandra Kanth N.** is currently employed at GE Digital, as a Data Engineering Specialist. He has completed his Bachelors in Computer Science and Engineering from R.V. College of Engineering, 2014-18. He started out as an Intern in the same company. He has published two technical papers and open sourced a deep learning framework for Face recognition. He loves to take part in Hackathons, Code-Sprints and has won two hackathons at national and international level. He is an avid football fan and loves to talk about latest technologies and advancements. He can be reached out at **e-mail id** - canfindck@gmail.com.